\documentclass[prd,showpacs,showkeys,floatfix,nofootinbib,eqsecnum,
                  preprint,12pt,tightenlines,fleqn]{revtex4} 

\usepackage{amsmath,amssymb,revsymb,graphicx,dcolumn}

\newcommand  {\version}{v6}

\newcommand{\rhovac}{\rho_\text{vac}}
\newcommand{\Pvac}  {P_\text{vac}}

\begin{document}

Phys. Rev. D 79, 063527 (2009)
\hfill arXiv:0811.4347 [gr-qc]  (\version)\newline\vspace*{3mm} 
\title[Gluonic vacuum, $q$--theory, and ...]
      {Gluonic vacuum, $q$--theory, and the cosmological constant\vspace*{5mm}}
\author{F.R. Klinkhamer}
\email{frans.klinkhamer@physik.uni-karlsruhe.de}
\affiliation{\mbox{Institute for Theoretical Physics, University of Karlsruhe (TH),}\\
76128 Karlsruhe, Germany}
\author{G.E. Volovik}
\email{volovik@boojum.hut.fi}
\affiliation{\mbox{Low Temperature Laboratory, Helsinki University of Technology,}\\
P.O. Box 5100, FIN-02015 HUT, Finland\\
and\\
\mbox{L.D. Landau Institute for Theoretical Physics, Russian Academy of Sciences,}\\
Kosygina 2, 119334 Moscow, Russia\\}

\begin{abstract}
\vspace*{.25\baselineskip}\noindent
In previous work,  $q$--theory was introduced to describe
the gravitating macroscopic behavior of a conserved microscopic variable $q$.
In this article, the gluon condensate of quantum chromodynamics is
considered in terms of $q$--theory.
The remnant vacuum energy density (i.e., cosmological constant)
of an expanding universe is estimated as $K_\text{QCD}^3/E_\text{Planck}^2$,
with string tension $K_\text{QCD} \approx \big(10^{2}\,\text{MeV}\big)^2$
and gravitational scale $E_\text{Planck} \approx 10^{19}\,\text{GeV}$.
The only input for this estimate is general relativity,
quantum chromodynamics, and the Hubble expansion of the present Universe.
\end{abstract}

\maketitle

\section{Introduction}
\label{sec:Introduction}

In a recent series of articles~\cite{KlinkhamerVolovik2008a,
KlinkhamerVolovik2008b,KlinkhamerVolovik2008c}, we explored
a new approach to the gravitational effects of vacuum energy density.
This approach starts from a conserved microscopic variable $q$,
whose statics and dynamics are studied on macroscopic scales.

The precise nature of $q$ is uncertain for the moment, but we have
presented at least one concrete example in terms of a four-form field $F$.
This $F$ field could be a part of the (unknown) fundamental theory of
elementary particle physics with an
energy scale given by $E_\text{Planck} \approx 10^{19}\,\text{GeV}$.

In this article, we do not contemplate ultrahigh energies but propose an
explicit realization of $q$ from well-established physics, namely, quantum
chromodynamics (QCD) with an energy scale of the order of $1\,\text{GeV}$.
That is, we find that $q$ can be identified as a particular gluon condensate
in the nonperturbative QCD vacuum.
Neglecting QCD effects, an $F$--type field~\cite{KlinkhamerVolovik2008b}
may still be required to reduce
the macroscopic vacuum energy density from a natural value of the order of
$(E_\text{Planck})^4$ to a value which is essentially zero .

With our general understanding of $q$--theory,
we can then investigate the gravitational
effects of the QCD vacuum. Most importantly, we find that the
dynamics of this gluon condensate in the nonequilibrium  context
of the expanding Universe may result in a \emph{nonzero} limiting
value of the vacuum energy density.
This remnant vacuum energy density may correspond to the inferred
cosmological constant responsible for the observed ``cosmic acceleration''
(cf. Ref.~\cite{Weinberg2008} and other references therein).

In order to be clear about the terminology,
we consider a time-dependent gravitating vacuum energy density $\rhovac(t)$
in an expanding Friedmann--Robertson--Walker (FRW)
universe [typically, $\rhovac(t)$ decreases
with cosmic time $t\,$] and \emph{define} the cosmological constant $\Lambda$
as the remnant  vacuum energy density in the limit of large cosmic times,
$\Lambda \equiv \lim_{t\to\infty}\rhovac(t)$.
This also implies that, while the vacuum energy density $\rho_\text{vac}$
may have changed with time, the equation-of-state parameter
$w_\text{vac} \equiv P_\text{vac}/\rho_\text{vac}$
has kept the value $-1$, at least for the type of theories considered here.
The possibility of having time-dependent $\rho_\text{vac}(t)$ and
constant $w_\text{vac}=-1$ may be an important lesson for observational cosmology.

Throughout this article, natural units are used with $c=\hbar=1$
and Newton's gravitational constant $G_\text{N}$ is shown explicitly.

\section{Gluon condensate}
\label{sec:Gluon-condensate}

The underlying theory of the strong interactions is nowadays believed to be
given by a particular non-Abelian gauge field theory called
quantum chromodynamics; see, e.g., Ref.~\cite{ChengLi1985}
and other references therein.
The  non-Abelian gauge group is  $SU(N_\text{c})$ and
the perturbative particle content of QCD is given by $N_\text{f}$ flavors of
quarks and $N_\text{c}^2-1=8$ types of gluons,
for $N_\text{c}=3$ colors of each quark flavor.
The nonperturbative particle content of QCD is given by the genuine
asymptotic states, that is, the baryons, the mesons, and possibly
the glueballs.

The crucial object for our discussion is the Yang--Mills field strength,
defined as
\begin{equation}
G_{\;\;\mu\nu}^{a}(x)=
\partial_\mu A_{\;\;\nu}^{a}(x) - \partial_\nu A_{\;\;\mu}^{a}(x)
 + f^{abc}\,A_{\;\;\mu}^{b}(x) A_{\;\;\nu}^{c}(x) \,,
\label{eq:defG}
\end{equation}
with spacetime indices $\mu,\nu$ ranging over $0$ to $3$,
Lie-algebra indices $a,b,c$ taking values from 1 to $N_\text{c}^2-1$,
repeated Lie-algebra indices $b,c$ being summed over, and
structure constants $f^{abc}$ corresponding
to the Lie algebra $\mathsf{su}(N_\text{c})$.
Note that the gauge coupling constant $g$ has been absorbed in the gauge
potential, so that $A_{\;\;\mu}^{a} =\text{O}(g^0)$ for an instanton
configuration and $A_{\;\;\mu}^{a} =\text{O}(g^1)$ for a perturbative
configuration with a few gluons.

Consider, now, the QCD gluon condensate
(see Ref.~\cite{ShifmanVainshteinZakharov1978} and other references therein).
It follows directly from gauge invariance that
\begin{equation}
\langle 0|\, G_{\;\;\mu\nu}^{a}(x)\, |0\rangle=0 \,,
\label{eq:VEVG}
\end{equation}
where the vacuum expectation value can, for example,
be obtained from a Euclidean path integral.
Equation \eqref{eq:VEVG} for the vacuum expectation value of a Yang--Mills
field of local support is a special case of Elitzur's
theorem~\cite{Elitzur1975}, which relies on the gauge noninvariance of the
Yang--Mills field strength and the gauge invariance of the path integral and
the vacuum state. As a further clarification of \eqref{eq:VEVG},
we state our explicit assumption
that the so-called Savvidy vacuum~\cite{Savvidy1977} is not realized,
i.e., that the vacuum expectation value of the average color magnetic field
is zero.

The vacuum expectation value of the quadratic expression can,
however, be nonzero:
\begin{eqnarray}
&&
\langle0|\,\frac{1}{4\pi^2}\,G_{\;\;\mu\nu}^{a}(x)\,G_{\;\;\rho\sigma}^{a}(x)\,|0\rangle
=                      
\frac{1}{12}\;q(x) \;\Big(g_{\mu\rho}(x)\, g_{\nu\sigma}(x) -g_{\mu\sigma}(x)\, g_{\nu\rho}(x)\Big)\,,
\label{eq:VEVG^2}
\end{eqnarray}
again with an implicit sum over the repeated Lie-algebra index $a$.
The explicit realization of the vacuum variable $q$
from Ref.~\cite{KlinkhamerVolovik2008a} is then
\begin{equation}
q(x)=\langle 0|\,  \frac{1}{4\pi^2}\, G^{a\;\mu\nu}(x)\,G_{\;\;\mu\nu}^{a}(x)\, |0\rangle\,,
\label{eq:q}
\end{equation}
which, with the chosen numerical factor,
is precisely equal to the Shifman--Vainshtein--Zakharov condensate
as determined from charmonium data $\big(q\approx 10^{-2}\,\text{GeV}^4\,\big)$.
On the theoretical side, note that the vacuum expectation value
\eqref{eq:q} is a properly renormalized quantity, which can, for example,
be obtained from a Euclidean path integral
calculated in the dilute-instanton-gas approximation
(see Sec.~6.7 of Ref.~\cite[(a)]{ShifmanVainshteinZakharov1978}
and other references therein).

The experimental value for $q$ is positive, even though
expression \eqref{eq:q} is not positive definite for a Lorentzian
spacetime metric. However, $q$ is manifestly positive definite for a Euclidean
spacetime metric, which is anyway needed to make sense of the path integrals
for instanton-type calculations.
Henceforth, we consider the vacuum variable $q$ to be non-negative.

Before we start our discussion of the gravitational effects of the
gluon condensate, we can already mention a side-product
of our investigation, namely, that the gluon condensate has a new
characteristic, the compressibility $\chi$. This will be
mentioned briefly in Secs.~\ref{sec:Effective-potential-for-q}
and \ref{sec:Lambda-from-self-sustained-gluonic-vacuum}
and Appendix~\ref{sec:App-A}, while the general discussion
of vacuum compressibility has been presented
in Ref.~\cite{KlinkhamerVolovik2008a}.

\section{Cosmological term for gravity}
\label{sec:Cosmological-term}

The goal of the present article is to explore certain gravitational effects
of the QCD gluon condensate over spacetime volumes very much larger
than those corresponding to the typical scales of QCD,
$\ell_\text{QCD}=c\,\tau_\text{QCD} \approx  1\,\text{fm} \equiv 10^{-15}\,\text{m}
\approx \hbar c/(200\:\text{MeV})$.
In the spirit of Ref.~\cite{KlinkhamerVolovik2008a},
we consider the following coarse-grained effective action:
\begin{eqnarray}
 S_\text{eff}[g,q]&=&S_\text{grav}[g] + S_\text{vac}[g,q]
 =                                
\int d^4x\, \sqrt{-g}\;\Big( \frac{1}{16\pi G_\text{N}}\;R[g] + \epsilon(q)\Big)\,,
\label{eq:VacuumAction}
\end{eqnarray}
where the pure-gravity action $S_\text{grav}$
is given by the standard Einstein--Hilbert
term with the Ricci curvature scalar $R=R[g]$ and the vacuum-energy-density
action $S_\text{vac}$ is determined by a general function $\epsilon(q)$ of
the vacuum variable $q$.
Here, $q$ is realized as the vacuum expectation value \eqref{eq:q}
but now averaged over a (Euclidean) spacetime volume of the order
of $(1\,\text{fm})^4$.

The energy-momentum tensor obtained by variation over $g^{\mu\nu}$
is then given by
\begin{equation}
T_{\mu\nu}=-\frac{2}{\sqrt{-g}}\: \frac{\delta S_\text{vac}}{\delta
g^{\mu\nu}}= \epsilon(q)\,  g_{\mu\nu} - 2\,  \frac{d\epsilon(q)}{d q}\,
\frac{\delta q}{\delta g^{\mu\nu}}
 \,.
\label{eq:em_tensor}
\end{equation}
Using \eqref{eq:VEVG^2} and \eqref{eq:q}, one obtains
\begin{equation}
\frac{\delta q}{\delta g^{\mu\nu}}
=2\left< \frac{1}{4\pi^2}\, G_{\;\;\mu\rho}^{a}G_{\;\;\nu\sigma}^{a}\right> \,g^{\rho\sigma}
=\frac{1}{2}\, q\, g_{\mu\nu}\,,
\label{eq:dq/dg}
\end{equation}
where the brackets of the expression in the middle
denote both the vacuum expectation value and
an average over a (Euclidean) spacetime volume of the order of $(1\,\text{fm})^4$
[the same expression can also be written in terms of the effective
fields \eqref{eq:master-gauge-field} introduced below].
As a result, \eqref{eq:em_tensor} produces a cosmological-constant-type
energy-momentum tensor for the Einstein field equation,
\begin{subequations}\label{eq:cosmological-term-T-rhovac}
\begin{eqnarray}
T_{\mu\nu}(q) &=&\rhovac(q)\, g_{\mu\nu}  \,,
\label{eq:cosmological-term-T}
\\[2mm]
\rhovac(q)    &=& \epsilon(q) - q\, \frac{d\epsilon(q)}{d q} \,,
\label{eq:cosmological-term-rhovac}
\end{eqnarray}
\end{subequations}
where the expression for $\rhovac(q)$ has precisely the structure argued on
thermodynamical grounds in Ref.~\cite{KlinkhamerVolovik2008a}.
The term \eqref{eq:cosmological-term-T} in a cosmological context
corresponds to a cosmic fluid with equation-of-state parameter $w_\text{vac}=-1$.

At this moment, it may be instructive to comment on
the difference between our approach and the one of nonlinear electrodynamics
as discussed in, e.g., Refs.~\cite{Novello-etal2004,Vollick2008}.
In our approach, the average energies of the magnetic and electric
field fluctuations in the vacuum are related
by $\left<\, |{\bf B}|^2\,\right>=-\left<\,|{\bf E}|^2\,\right>=q/4$,
as follows from \eqref{eq:VEVG^2}. (The negative value of
the average energy of the electric field is obtained after renormalization
of the divergent energy of the quantum fluctuations.)
But, in the approach of Refs.~\cite{Novello-etal2004,Vollick2008},
both quantities are non-negative, $\left<\,|{\bf B}|^2\,\right>\geq 0$
and $\left<\,|{\bf E}|^2\,\right>\geq 0$,
so that generically the energy-momentum tensor from the electromagnetic field
does not correspond to a cosmological-constant-type
term \eqref{eq:cosmological-term-T}.

\section{Equation for $\boldsymbol{q}$}
\label{sec:Equation-for-q}

The equation of motion for $q$ can be obtained by
averaging the effective Yang--Mills equation.
We proceed by the introduction of a  ``master gauge field''
(denoted by a bar), with the following properties
\begin{subequations}\label{eq:master-gauge-field}
\begin{eqnarray}
q &=& 1/(4\pi^2)\; \overline{G}^{\,a\;\mu\nu}\,\overline{G}_{\;\;\mu\nu}^{\,a}\,,
\label{eq:master-gauge-field-q}\\[1mm]
\overline{G}_{\;\;\mu\nu}^{\,a}
&=&
\partial_\mu \overline{A}_{\;\;\nu}^{\,a} - \partial_\nu \overline{A}_{\;\;\mu}^{\,a}
+ f^{abc}\,\overline{A}_{\;\;\mu}^{\,b} \overline{A}_{\;\;\nu}^{\,c}  \,.
\label{eq:master-gauge-field-Gbar}
\end{eqnarray}\end{subequations}
Physically, the idea is that the classical master field describes the gluon
condensate and allows for variations over spatial and temporal scales which
are large compared to the microscopic scales of QCD. Theoretically, such a
master field is known to exist in the large--$N_\text{c}$ limit; cf.
Ref.~\cite{GreensiteHalpern1983}.
In a follow-up article, the gradient-expansion
method will be used, which allows us to get
more explicit results [in this method, the vacuum order parameter $q(x)$
is considered to be a slow (hydrodynamic) variable with a length scale
of inhomogeneities large compared to the QCD length scale].

Now, start from the variation of \eqref{eq:VacuumAction} with respect to
$\overline{A}_\mu(x)\equiv \overline{A}_\mu^{\,a}(x)\, T^{a}\,$,
where the $T^{a}$ are the anti-Hermitian generators of the Lie algebra
$\mathsf{su}(N_\text{c})$. The variational principle then gives the
following field equation:
\begin{equation}
\overline{D}_\mu
\left(\frac{d\epsilon(q)}{d q}\;\overline{G}^{\,a\;\mu\nu}\, T^{a} \right) =0 \,,
\label{eq:Maxwell}
\end{equation}
where $q$ appearing in the function $\epsilon^\prime(q)\equiv d\epsilon/dq$
stands for the $\overline{G}^{\,2}$ expression \eqref{eq:master-gauge-field-q} and
$\overline{D}_\mu$ denotes the covariant derivative with respect to
general coordinate transformations
[using the standard affine connection $\Gamma_{\mu\nu}^{\lambda}(x)$]
and non-Abelian gauge transformations
[using the master gauge field $\overline{A}_{\;\;\mu}^{\,a}(x)$].

Next, contract \eqref{eq:Maxwell} with
$\overline{G}_{\kappa\nu}\equiv \overline{G}_{\;\;\kappa\nu}^{\,a}\, T^{a}$,
multiply by $1/(4\pi^2)$, and take the trace (with normalization factor $-2$):
\begin{equation}
-2\,\text{tr}\left[\;\frac{1}{4\pi^2}\;\overline{G}_{\kappa\nu}\,
           \overline{D}_\mu \left(\frac{d\epsilon(q)}{d q}\;\overline{G}^{\,\mu\nu}
\right)\right]=0 \,.
\label{eq:AverageMaxwell}
\end{equation}
Using
\begin{eqnarray}
-2\,\text{tr}\left[\overline{G}_{\kappa\nu}\, \overline{G}^{\,\mu\nu}\right]
&\equiv&
\overline{G}_{\;\;\kappa\nu}^{a}\, \overline{G}^{\,a\;\mu\nu}
= q\,\pi^2 ~ \delta_\kappa^{\;\;\mu}  \,,
\label{eq:q-as-trace}
\end{eqnarray}
one obtains
\begin{equation}
q\,\overline{D}_\kappa \left(\frac{d\epsilon(q)}{d q}  \right)  =
4\; \frac{d\epsilon(q)}{d q}\;2\,
\text{tr}\left[\frac{1}{4\pi^2}\;\overline{G}_{\kappa\nu}\,
           \overline{D}_\mu \overline{G}^{\,\mu\nu}\right] \,.
\label{eq:eq_for_q}
\end{equation}

At this moment, we can proceed in two directions.
The first direction assumes that the physical situation is
such that the master field satisfies the standard Yang--Mills equation,
$\overline{D}_\mu \overline{G}^{\,\mu\nu}=0$. Then, the following result holds:
\begin{eqnarray}
\text{tr}\left[\overline{G}_{\kappa\nu}\,
\overline{D}_\mu \overline{G}^{\,\mu\nu}\right]&=&0 \,,
\label{eq:GDG-zero}
\end{eqnarray}
which nullifies the right-hand side of \eqref{eq:eq_for_q}.

Equation \eqref{eq:GDG-zero} for the case of Minkowski spacetime
can also be argued as follows.
Take for granted that the nonperturbative QCD vacuum
over Minkowski spacetime does not break spacetime translation invariance
and also does not break any of the discrete symmetries of
charge conjugation ($\mathsf{C}$), parity reflection ($\mathsf{P}$), or
time-reversal ($\mathsf{T}$).
(The implicit assumption is that the so-called $\theta$
parameter vanishes; cf. Ref.~\cite{ChengLi1985}.)
Then, it follows that the $\kappa=0$ component  and the
$\kappa=1,2,3$ components of the left-hand side of \eqref{eq:GDG-zero}
vanish by, respectively, the $\mathsf{T}$ and $\mathsf{P}$
invariance of the Minkowski-spacetime QCD vacuum.

As $q$ is a gauge-invariant scalar, the covariant derivative
$\overline{D}_\kappa$ on the left-hand side of \eqref{eq:eq_for_q}
equals the standard gradient $\partial_\kappa$ and the solution
of \eqref{eq:eq_for_q} using \eqref{eq:GDG-zero} is simply
\begin{equation}
 \frac{d\epsilon(q)}{d q}   = \mu \,,
\label{eq:ChemPot}
\end{equation}
where $\mu$ is an integration constant.
[It will be shown in a forthcoming publication that
\eqref{eq:ChemPot} also follows from the gradient expansion up
to the first-order (linear) term in $\partial_\kappa q$.]
Result \eqref{eq:ChemPot} demonstrates that the density $q$
of the gluon condensate is a conserved quantity and that $\mu$ from
\eqref{eq:ChemPot} plays the role of the corresponding chemical potential.
The physical situation corresponds, therefore, to that of an equilibrium
state of the vacuum.

The second direction considers a physical situation  with
additional higher-derivative terms
contributing to the equation of motion for the master field,
so that $\overline{D}_\mu \overline{G}^{\,\mu\nu}$ need not vanish in general.
The equation of motion \eqref{eq:eq_for_q} can then be rewritten as
\begin{subequations}\label{eq:mubar-eq-def}
\begin{eqnarray}
\hspace*{-5mm}
\partial_\kappa\,  \overline{\mu} &=& -4 \, \overline{\mu}\;\;
\frac{\Big(-2\,\text{tr}\left[1/(4\pi^2)\;\overline{G}_{\kappa\nu}\,
           \overline{D}_\mu \overline{G}^{\mu\nu}\right]\Big)}
{\Big(-2\,\text{tr}\left[1/(4\pi^2)\;\overline{G}_{\mu\nu}\,
            \overline{G}^{\mu\nu}\right]\Big)}\;,
\label{eq:eq_for_mubar}\\[1mm]
\hspace*{-5mm}
\overline{\mu} &\equiv& \frac{d\epsilon(q)}{d q} \,,
\label{eq:def-mubar}
\end{eqnarray}
\end{subequations}
where $\overline{\mu}(x)$ is an effective gauge-invariant scalar field
and the denominator on the right-hand side of \eqref{eq:eq_for_mubar}
is precisely equal to $q$ according to
\eqref{eq:q-as-trace} with its spacetime indices contracted.
The dynamical equation \eqref{eq:eq_for_mubar}
will be used in Sec.~\ref{sec:Lambda-from-self-sustained-gluonic-vacuum}
and Appendix~\ref{sec:App-A} to estimate the remnant vacuum energy density
for the present ($\mathsf{T}$--noninvariant) Universe.

\section{Effective potential for $\boldsymbol{q}$}
\label{sec:Effective-potential-for-q}

In the simplest approach, the effective potential for $q$ is determined by
asymptotic freedom~\cite{GrossWilczekPolitzer1973}
and the conformal anomaly~\cite{NielsenDuncan-etal1977}
evaluated at one loop:
\begin{subequations}\label{eq:EffPot}
\begin{eqnarray}
\epsilon(q)   &=&\epsilon_0+ b_{1}\, q \, \ln \frac{q}{q_c}  \,,
\label{eq:EffPot-epsilon}\\[1mm]
b_{1}&=&
\frac{1}{32} \left(\frac{11}{3}\,N_\text{c} - \frac{2}{3}\,N_\text{f} \right)\,,
\label{eq:EffPot-b1}
\end{eqnarray}
\end{subequations}
with number of colors $N_\text{c} =3$
and number of light-quark flavors $N_\text{f} =2$ for QCD at low energies
[recall that $q$ as defined by \eqref{eq:q} contains an explicit factor $1/(4\pi^2)$].

With this choice for the effective potential $\epsilon(q)$, \eqref{eq:ChemPot}
gives the following expressions for the gluon-condensate charge $q$,
the macroscopic vacuum energy density $\rhovac$,
and the energy-momentum-tensor trace $T_\rho^{\;\;\rho}$
as a function of the chemical potential $\mu$:
\begin{subequations}\label{eq:q-rhovac-traceTmunu(mu)}
\begin{eqnarray}
q(\mu)&=&  q_c \, \exp\big(\mu/b_{1} -1 \big)
\,,
\label{eq:q(mu)}
\\[1mm]
\rhovac(\mu)&=&\epsilon\big(q(\mu)\big)   -\mu\, q(\mu)
             = \epsilon_0-    b_{1}\, q(\mu)  \,,
\label{eq:rhovac(mu)}
\\[1mm]
T_\rho^{\;\;\rho}(\mu) &=&4\,\epsilon_0 -4\, b_{1}\, q(\mu)   \,.
\label{eq:traceTmunu(mu)}
\end{eqnarray}
\end{subequations}
The second term on the right-hand side of \eqref{eq:traceTmunu(mu)} corresponds
to the conformal anomaly, as discussed in, e.g., Ref.~\cite{Zhitnitsky2007},
where $\mu$ is the chemical potential of baryons and reflects the conservation
of baryonic charge.

Here, $\mu$ is the chemical potential that
characterizes the vacuum state and
reflects conservation of the vacuum charge $q$.
Moreover, $\mu$ becomes a ``running coupling constant,''
\begin{equation}
\mu= b_{1}\,\big(1+ \ln (q/q_c) \big)\,,
\label{eq:running_coupling}
\end{equation}
as follows from \eqref{eq:q(mu)}. The gluonic vacuum is stable, since the vacuum
compressibility~\cite{KlinkhamerVolovik2008a} is positive for $b_{1}>0$
and $q>0$:
\begin{equation}
\chi= \left(q^2\, \frac{d^2\epsilon}{dq^2} \right)^{-1}
    =      \left( b_{1}\, q \right)^{-1}  \,,
\label{eq:compressibility}
\end{equation}
as will be discussed further in the next section.

\section{$\boldsymbol{\Lambda}$ from a self-sustained gluonic vacuum}
\label{sec:Lambda-from-self-sustained-gluonic-vacuum}

Given that $q$ from \eqref{eq:q} and $b_{1}$ from \eqref{eq:EffPot-b1} are
non-negative for low-energy QCD,
the vacuum energy density $\rhovac(\mu)$ in \eqref{eq:rhovac(mu)}
can be nullified if $\epsilon_0>0$. In this case,  the self-sustained vacuum
is given by
\begin{subequations}\label{eq:SSvacuum}
\begin{eqnarray}
\mu_0&=&b_{1}\,\Big( 1-\ln b_{1} + \ln\big(\epsilon_0/q_c \big)\Big)\,,
\label{eq:SSvacuum-mu0} \\[1mm]
q(\mu_0)&\equiv& q_0   =  \epsilon_0/b_{1}\,,
\label{eq:SSvacuum-q0}
 \\[1mm]
\rhovac(\mu_0)&=&-P_\text{vac}(\mu_0)=0 \,,
\label{eq:SSvacuum-rhovac0}
 \\[1mm]
T_\rho^{\;\;\rho} (\mu_0)&=&0 \,,
\label{eq:SSvacuum-traceTmunu(mu0)}
\end{eqnarray}\end{subequations}
where the result for the vacuum pressure in \eqref{eq:SSvacuum-rhovac0}
follows from the general energy-momentum tensor \eqref{eq:cosmological-term-T}.
Recall that a self-sustained vacuum~\cite{KlinkhamerVolovik2008a}
can exist as an equilibrium state at zero external pressure $P_\text{ext}$,
with pressure equilibrium giving $P_\text{vac}=P_\text{ext}=0$.
The particular gluon-condensate vacuum discussed here has a
vacuum compressibility \eqref{eq:compressibility} given by
$\chi_0 \equiv \chi(q_0) = 1/\epsilon_0 >0$, according to \eqref{eq:SSvacuum-q0}.

For the case $\epsilon_0<0$ and with $q>0$,
the energy density \eqref{eq:rhovac(mu)} can only be nullified
if $b_{1}<0$, which holds for an Abelian gauge field theory such as QED.
The vacuum would, however, be unstable, since
the vacuum compressibility \eqref{eq:compressibility}
would be negative for negative $b_{1}$.
In addition, it is far from obvious
that a nonzero vacuum expectation value \eqref{eq:q} for $q$
can arise in an Abelian gauge field theory.
In short, a stable self-sustained vacuum can be realized by a non-Abelian
gauge field theory with $\epsilon_0>0$ but not by an Abelian gauge field
theory.\footnote{The quantity $\epsilon_0$ may, in principle,
come as the response of (or reaction from)
the deep vacuum at the Planck-energy scale, which is slightly adjusted
to compensate the energy of the gluon condensate, as discussed in Sec.~II of
Ref.~\cite{KlinkhamerVolovik2008a} for the case of a scalar condensate.
For the present discussion, $\epsilon_0$ is simply assumed to be
positive. As mentioned above,
$\epsilon_0$ then corresponds to the inverse vacuum compressibility.}

The quantities $q(\mu_0)$ and $q_c$ are determined by the
characteristic QCD energy scale $\Lambda_\text{QCD}$
from the asymptotic-freedom behavior~\cite{GrossWilczekPolitzer1973}
of the $SU(3)$ gauge coupling constant,
\begin{equation}
q(\mu_0) \sim q_c
         \sim \Lambda_\text{QCD}^4
         \approx \big(200\,\text{MeV}\big)^4 \,,
\label{eq:scales}
\end{equation}
with $\epsilon_0\sim b_1\,\Lambda_\text{QCD}^4$ from \eqref{eq:SSvacuum-q0}.
Still, the macroscopic energy density of the self-sustained vacuum that
enters the Einstein equation as a cosmological constant is not given by
$\epsilon(q_0) \approx \epsilon_0 = \text{O}\big( 10^{33}\,\text{eV}^4 \big)$
but is strictly zero, $\Lambda=\rhovac(\mu_0)=0$, according
to \eqref{eq:cosmological-term-T-rhovac}, \eqref{eq:ChemPot},
and \eqref{eq:SSvacuum-rhovac0}.

A nonzero value of $\Lambda=\rhovac(\mu)$ may appear for a perturbed vacuum
with $\mu\ne \mu_0$. Specifically, the vacuum energy density induced by the
expansion of the Universe
can be expected to be nonzero (cf. Sec.~\ref{sec:Equation-for-q}).
Based on the heuristic discussion in Appendix~\ref{sec:App-A},
we suggest the following behavior:
\begin{equation}
\rhovac \sim f\,|H|\,\Lambda_\text{QCD}^3\,,
\label{eq:Induced-rhovac}
\end{equation}
with Hubble parameter $H\equiv (da/dt)/a > 0$ for an expanding universe
and a numerical factor $f \geq 0$.
It has indeed been argued~\cite{Schutzhold2002} on general grounds that
the linear $H$ term of \eqref{eq:Induced-rhovac}
may arise from the nonperturbative QCD interactions that
anomalously break the scale invariance of the massless classical theory.
The potential importance of the QCD vacuum for cosmological horizons
has also been emphasized in Ref.~\cite{Bjorken2004}.

According to our present understanding (see, e.g., Ref.~\cite{Weinberg2008}),
the Universe evolved from an early radiation/matter-dominated
phase [$H(t) \sim 1/t$] to a late vacuum-dominated phase [$H(t) \sim \text{const.}$].
The crossover will be discussed further in the next section, but, here,
only the asymptotic behavior ($t \to \infty$) will be considered.

For a stationary de-Sitter universe, result \eqref{eq:Induced-rhovac}
can be written as
\begin{equation}
\Lambda = f\,H_\text{deS}\,\Lambda_\text{QCD}^3 \,,
\label{eq:InducedLambda}
\end{equation}
with $H_\text{deS}>0$  the Hubble constant (time-independent Hubble parameter)
of de-Sitter spacetime and neglecting higher-order terms
such as $H_\text{deS}^2\,\Lambda_\text{QCD}^2$.
In addition, the standard Friedmann equation gives for a de-Sitter universe
\begin{equation}
H_\text{deS}^2=(8\pi/3)\;\Lambda/E_\text{Planck}^2\,,
\label{eq:deSitter}
\end{equation}
with $E_\text{Planck} \equiv \sqrt{\hbar\, c^5/G_\text{N}}
\approx 1.22 \times 10^{28}\,\text{eV}$.
Eliminating $H_\text{deS}$ from the last two equations, one obtains the
following estimate of the cosmological constant (remnant vacuum energy density):
\begin{equation}
\Lambda =  (8\pi/3)\;f^2\,\Lambda_\text{QCD}^6/E_\text{Planck}^2\,,
\label{eq:InducedLambda2}
\end{equation}
where the numerical constant $f^2$ remains to be determined.

As the QCD scale parameter $\Lambda_\text{QCD}$
of the $SU(3)$ gauge coupling constant is
renormalization-scheme dependent, it may be more appropriate, conceptually,
to give the cosmological constant in terms of a directly measurable quantity.
Specifically, we take the string tension
$K_\text{QCD} \equiv 1/\big(2\pi \alpha^\prime\big)\approx \big(400\,\text{MeV}\big)^2$
from the measured Regge slope $\alpha^\prime$ of meson resonances~\cite{ChengLi1985}
or from numerical calculations of lattice gauge theory~\cite{Bali2000}
combined with other experimental data to fix the absolute length scale.
Setting $\Lambda_\text{QCD}^2 \approx K_\text{QCD}/4$
and $f^2 \approx (24/\pi)\,k_\Lambda$ in \eqref{eq:InducedLambda2},
the final expression for the cosmological constant reads
\begin{eqnarray}
\Lambda      &=& k_\Lambda \,K_\text{QCD}^3/E_\text{Planck}^2
               \approx                               
       \Big(3 \times 10^{-3}\,\text{eV}\Big)^4       
       \left(\frac{k_\Lambda}{2 \times 10^{-6}}\right)
       \left(\frac{K_\text{QCD}}{\big(400\,\text{MeV}\big)^2}\right)^{3},
\label{eq:InducedLambda3}
\end{eqnarray}
where the numerical constant $k_\Lambda$ remains to be determined
(the experimental results to be discussed shortly suggest a value
of the order of $10^{-6}$).
Result \eqref{eq:InducedLambda3} can also be written
as $\Lambda  \sim  \big(G_\text{N}/c^5\big)\,K_\text{QCD}^3/\hbar$,
in order to emphasize that
the result relies only on classical general relativity
and quantum chromodynamics (the string tension $K_\text{QCD}$ has
the dimension of energy over length).

The suggestion, then, is that the vacuum of the presently observed Universe is not
relaxing to the absolute equilibrium state
\eqref{eq:SSvacuum} but to the de-Sitter equilibrium state with nonzero
cosmological constant \eqref{eq:InducedLambda2} or
equivalently  \eqref{eq:InducedLambda3}.

Since the proton mass $m_p \approx 938\,\text{MeV}$ is now known
to come mostly from the gluon dynamics, $m_p\sim \Lambda_\text{QCD}$,
estimate \eqref{eq:InducedLambda2} corresponds to Zeldovich's
original suggestion~\cite{Zeldovich1967} for the cosmological constant
in terms of the proton mass,
$\Lambda_\text{Zeldovich}\sim m_p^6/E_\text{Planck}^2$.
Numerically, one has $m_p^6\gg \Lambda_\text{QCD}^6$
and Zeldovich's expression gives a value for $\Lambda$
several orders of magnitude larger than \eqref{eq:InducedLambda2},
which is closer to the observed value but still somewhat too large for $f=1$.

The numerical agreement between the theoretical estimate
\eqref{eq:InducedLambda2} or \eqref{eq:InducedLambda3}
and the experimental value~\cite{Weinberg2008,Komatsu2008,Vikhlinin-etal2008}
of approximately $\big(2\,\text{meV}\big)^4$
is improved by having a reduction factor
$k_\Lambda =\text{O}\big(10^{-6}\big)$ in \eqref{eq:InducedLambda3}.
The corresponding factor $f=\text{O}\big(10^{-3}\big)$ in
\eqref{eq:InducedLambda2} traces back to \eqref{eq:Induced-rhovac} and
depends on the evolution of the gluon condensate as the Universe cools from
$T \approx 200\,\text{MeV}$ to the present temperature $T \approx 3\,\text{K}$;
see Appendix~\ref{sec:App-A} for further details.

\section{Other cosmological constant problems}
\label{sec:Other-CC-problems}

In the previous section, we have made an attempt
to use QCD for the following three cosmological constant problems
(cf. Refs.~\cite{KlinkhamerVolovik2008a,KlinkhamerVolovik2008b,
KlinkhamerVolovik2008c,Weinberg2008} and other references therein):
\begin{itemize}
\item[(i)]  why is the cosmological constant $\Lambda$ not catastrophically large?
\item[(ii)] why does not $\Lambda$ vanish exactly?
\item[(iii)] what physical mechanism sets the scale of $\Lambda$?
\end{itemize}
 From the $q$--theory  approach to QCD, we have found that
the Universe asymptotically approaches a stationary de-Sitter
phase with a cosmological constant $\Lambda$
given by \eqref{eq:InducedLambda2} or equivalently \eqref{eq:InducedLambda3},
which suggests a partial solution to the above
three problems.\footnote{As mentioned in Sec.~\ref{sec:Introduction},
non--QCD contributions to the vacuum energy density are perhaps canceled
by the self-adjustment of another $q$--type field such as the
4--form field $F$ considered in Ref.~\cite{KlinkhamerVolovik2008b}
or by an entirely different mechanism.}

However, not all cosmological constant problems have been addressed,
let alone solved completely.
There remain, for example, the following two questions:
\begin{itemize}
\item[(iv)]
at which moment in time, $t=t_\text{cross}$, does the vacuum energy density
start to dominate over the cold-dark-matter energy density?
\item[(v)]
why do galaxies and stars exist at times relatively close to $t_\text{cross}$?
\end{itemize}

Question (iv) perhaps has a simple answer in our approach.
The crossover from the cold-dark-matter-dominated Universe to
the gluon-condensate-dominated Universe occurs
when the cold-dark-matter energy density drops below the vacuum energy density.
For a flat matter-dominated FRW universe
with Hubble expansion parameter $H \approx (2/3)\,1/t$,
the cold-dark-matter energy density evolves as $\rho_{\text{CDM}}(t)
\approx (3/8\pi)\,(4/9)\,E_\text{Planck}^2/t^2$, with
numerical factors of order unity displayed.\footnote{In first
approximation, the energy transfer from vacuum to cold dark matter
can be neglected for $t \ll t_\text{cross}$, as long as
the vacuum energy density is given by \eqref{eq:Induced-rhovac}
also for a time-dependent Hubble parameter $H$. The question remains
as to the precise nature of the energy exchange between vacuum and
matter~\cite{Klinkhamer2008}.}  Asymptotically ($t \gg t_\text{cross}$),
the constant vacuum energy density is given by \eqref{eq:InducedLambda2}.
The resulting crossover time can, therefore, be estimated as
\begin{equation}
\hspace*{-6mm}
t_\text{cross} \approx \big(4\pi\big)^{-1}\;f^{-1}\,
                        E_\text{Planck}^2/\Lambda_\text{QCD}^3 \,.
\label{eq:t-cross-tmp}
\end{equation}
In terms of the string tension $K_\text{QCD}\approx 4\,\Lambda_\text{QCD}^2$
and the numerical constant $k_\Lambda \approx (\pi/24)\,f^2$,
the crossover time \eqref{eq:t-cross-tmp} becomes
\begin{eqnarray}
\hspace*{-6mm}
t_\text{cross} &\approx& \big(6\pi\,k_\Lambda\big)^{-1/2}\;
                          E_\text{Planck}^2 \,K_\text{QCD}^{-3/2}
                 \approx                                
                       2 \times 10^{17}\,\text{s}   
                       \left(\frac{2 \times 10^{-6}}{k_\Lambda}\right)^{1/2}
                       \left(\frac{\big(400\,\text{MeV}\big)^2}
                       {K_\text{QCD}}\right)^{3/2},
\label{eq:t-cross}
\end{eqnarray}
where a value for $k_\Lambda$ of the order of $10^{-6}$ is indicated
by the comparison of \eqref{eq:InducedLambda3} with the measured vacuum
energy density, as discussed in
Sec.~\ref{sec:Lambda-from-self-sustained-gluonic-vacuum}.

The value for $t_\text{cross}$ from \eqref{eq:t-cross} is of the order of
and even just under the observed value for the age of the present Universe,
$t_0 \approx 14 \,\text{Gyr} \approx 4\times 10^{17}\,\text{s}\,$,
as determined from the data compiled in
Refs.~\cite{Komatsu2008,Vikhlinin-etal2008}.
The corresponding redshift $z_\text{cross}=\text{O}(1)$
agrees with the results indicated by deep supernovae observations,
such as those reported in Ref.~\cite{Riess2007}.\footnote{After
the completion of an earlier version of the present article,
we became aware of recent work~\cite{Carneiro-etal2006} on
the evolution of an expanding FRW universe with
a hypothetical decaying vacuum energy density proportional
to the Hubble parameter.}

Question (v) remains unanswered for the moment, but the answer
could also be related to QCD, possibly via the mass and baryon number
of the proton.\footnote{It is known~\cite{Dicke1961} that the
lifetime $t_\star$ of a main-sequence star
is, to a large extent, determined by the proton mass and Newton's constant:
$t_\star\sim E_\text{Planck}^2/m_p^3$,
with an additional numerical factor of the order of $10^4$ on the right-hand side
for a solar-mass star. This numerical factor contains, however,
the fine-structure constant $\alpha$ and the proton-electron mass ratio $m_p/m_e$,
so that non--QCD physics enters the estimate. The estimate for $t_\star$ is,
therefore, suggestive but inconclusive from our point of view.}

\section{Conclusion}
\label{sec:Conclusion}

In this article, we have described the nonperturbative QCD vacuum
in terms of $q$--theory, where $q$ is identified with the particular
gluon condensate \eqref{eq:q}.
A crucial role is played by the QCD trace anomaly~\cite{NielsenDuncan-etal1977},
whose potential relevance to the cosmological constant problem has previously
been emphasized in, e.g., Ref.~\cite{Schutzhold2002}
(see also Ref.~\cite{LabunRafelski2008} for a discussion in the context of QED).

The static equilibrium $q$--theory gives a gravitating vacuum energy density
which is exactly zero, $\rho_{\text{vac}} = 0$, according to
\eqref{eq:SSvacuum}.  But in a nonstatic
situation (e.g., that of the expanding Universe),
the gluon condensate is perturbed and a nonzero
gravitating vacuum energy density results, $\rho_{\text{vac}}\ne 0$.
The theoretical value for the remnant vacuum energy density
(cosmological constant) is estimated to be
given by \eqref{eq:InducedLambda3} and appears to be of the order of the
experimental value~\cite{Weinberg2008,Komatsu2008,Vikhlinin-etal2008}.

However, a reliable calculation of the present vacuum energy density
$\rho_{\text{vac}}$ will require a detailed study of the
gluon-condensate dynamics in an expanding universe. Even though that
study has barely started and many questions remain,
it is remarkable and encouraging that an
explanation of the so-called ``dark energy'' can perhaps be found
in known physics, classical general relativity and quantum chromodynamics.

\section*{\hspace*{-4.5mm}ACKNOWLEDGMENTS}

It is a pleasure to thank A.A. Starobinsky and the referee for helpful comments.
GEV is supported in part by the Russian Foundation for Basic Research
(Grant No. 06--02--16002--a)
and the Khalatnikov--Starobinsky leading scientific school (Grant No. 4899.2008.2).

\begin{appendix}

\section{Remnant vacuum energy density from QCD}
\label{sec:App-A}

In this appendix, we give a heuristic derivation of expression
\eqref{eq:Induced-rhovac} for the remnant vacuum energy density from
the gluon condensate of quantum chromodynamics in an expanding universe.
The main idea is that nonanalytic behavior of the gravitating
vacuum energy density as a function of the Hubble parameter
$H(t)\equiv (da/dt)/a$ may come from the singularity of the gluon propagator
in the infrared.

In order to see how this may happen, it is convenient to use the Gribov
picture of confinement~\cite{Gribov1978,Zwanziger1997,Burgio-etal2008}.
In the Coulomb gauge, the effective gluon mass would then depend on the
three-momentum $\mathbf{k}$ and would
increase in the infrared limit $k\equiv |{\bf k}|\rightarrow 0$ as
 \begin{equation}
m(\mathbf{ k}) \sim \Lambda_\text{QCD}^2/k\,.
\label{eq:Gribov_mass}
\end{equation}
Such a momentum-dependent mass would come from the instantaneous Coulomb
interaction between the color charges of the gluons, i.e.,
from the interaction potential $U(\mathbf{ r})\propto 1/r$ in coordinate space
or $U_\mathbf{ k}\propto 1/k^2$ in momentum space.

Here, we consider the possible effects from a more singular behavior
of $m(\mathbf{ k})$ at extremely small $k$.
Assuming a linear confinement potential between gluons
$U(\mathbf{ r})\propto \Lambda_\text{QCD}^2 \,r$,
with Fourier transform $U_\mathbf{ k}\propto \Lambda_\text{QCD}^5/k^4$,
we have the following behavior of the effective gluon mass in the
extreme infrared region:
 \begin{equation}
m(\mathbf{ k}) \sim \Lambda_\text{QCD}^3/k^2\,.
\label{eq:Gribov_mass_extension}
\end{equation}
Recall that, on the one hand, the Richardson potential~\cite{Richardson1979}
with a $1/k^2$ behavior of the effective gluon mass gives a reasonable description
of heavy-quark systems and that, on the other hand, a large--$N_\text{c}$
master field has been suggested~\cite{GreensiteHalpern1983},
which gives this very same potential for color sources in arbitrary
nontrivial representations of $SU(N_\text{c})$.
Current lattice-gauge-theory simulations~\cite{Burgio-etal2008} appear to
support the behavior \eqref{eq:Gribov_mass}, but are by necessity limited to
rather small volumes of the order of $(1\;\text{fm})^3$.
The conjectured behavior \eqref{eq:Gribov_mass_extension} would hold over length
scales $L$ larger than $1\;\text{fm}$ (perhaps $L \gtrsim 10\;\text{fm}$) and may
provide an incentive to push the pure-gauge lattice simulations to their limit.

In the cosmological context, a natural infrared cutoff for the divergent
gluon mass is provided by the Hubble expansion,
\begin{equation}
m(\mathbf{ k},H) \sim \Lambda_\text{QCD}^3\big/\big(k^2+H^2 \big)\,.
\label{eq:mass_H}
\end{equation}
The contribution of the Hubble expansion
to the vacuum energy density can be estimated by using,
for example, the zero-point energy of the gluon field. For the
FRW universe (or, more specifically, the de-Sitter universe),
the estimated contribution of zero-point energies
from \eqref{eq:mass_H} is
\begin{eqnarray}
\rhovac(H)
&\sim&
\frac{N_c^2-1}{2}\int \frac{d^3k}{(2\pi)^3}\,
\Big(m(\mathbf{ k},H) -m(\mathbf{ k},0)\Big)
\sim                       
 -\frac{N_c^2-1}{8\pi}\;|H| \;\Lambda_\text{QCD}^3 \,,
\label{eq:zero_point_frw}
\end{eqnarray}
where the factor $N_c^2-1$ counts the number of gluons in a pure $SU(N_c)$
Yang--Mills theory.
As argued in the main text, the vanishing of the gravitating
vacuum energy density $\rhovac$ in
Minkowski spacetime ($H=0$) would be due to the self-adjustment
of a $q$--type variable.

Even though \eqref{eq:zero_point_frw} has the wrong sign
(cosmology~\cite{Komatsu2008,Vikhlinin-etal2008,Riess2007}
suggests $\rhovac=-\Pvac>0$), the important point
is to have found that a term of order $|H| \;\Lambda_\text{QCD}^3$
can arise at all. The contributions of the fermionic quarks,
which have not been considered up till now,
may, in principle, reverse the overall sign of \eqref{eq:zero_point_frw}.
In any case, the zero-point-energy estimates,
which are applicable to equilibrium vacua, have only heuristic value
if the dynamics of the nonequilibrium vacuum is considered:
these estimates may give the correct order of magnitude
but not the exact number or even the sign.

Turning to the dynamics,
the infrared behavior of QCD in \eqref{eq:Gribov_mass_extension}
induces nonanalytic higher-order derivative terms in the gradient expansion
mentioned in Sec.~\ref{sec:Equation-for-q}.
The singular infrared behavior also leads to nonanalytic higher-derivative
terms in the master-field equation relevant to right-hand side
of \eqref{eq:eq_for_mubar}. This could, for example, give\footnote{The quartic root of
the differential operator $\mathcal{D}\equiv \overline{\square}^{\;2}$
in \eqref{eq:master-field-equation} can be defined as
$\mathcal{D}^{1/4} \equiv \lim_{\eta \downarrow 0}\;(2\sqrt{2}/\pi)\;\mathcal{D}\,
                \int_{\eta}^{\infty} dk\, \big( k^4 + \mathcal{D} \big)^{-1}$.
Note that the eigenvalues of $\mathcal{D}$ are non-negative, also
for a spacetime metric with Lorentzian signature.}
\begin{equation}
\overline{G}_{\,\kappa\nu}\,\overline{D}_\mu\,\overline{G}^{\,\mu\nu}
= c_1\,\tau_\text{QCD}\,
\overline{D}_\kappa\,\big(\overline{\square}\,\overline{\square}\,\big)^{1/4}\;
\Big(\overline{G}_{\,\mu\nu}\,\overline{G}^{\,\mu\nu}\Big)+\cdots \,,
\label{eq:master-field-equation}
\end{equation}
with a numerical coefficient $c_1$,
the microscopic time scale $\tau_\text{QCD} \sim 1/\Lambda_\text{QCD}$, and the
invariant d'Alembertian $\overline{\square}$ defined in terms of the master gauge
field $\overline{A}_{\mu}(x)$ and the standard affine connection
$\Gamma^{\lambda}_{\mu\nu}(x)$ from the metric $g_{\mu\nu}(x)$
and its inverse~\cite{Weinberg2008}.

For a flat FRW universe with a
time-dependent homogenous master field \eqref{eq:master-gauge-field-Gbar}
and corresponding scalar field $\overline{\mu}(t)$ from
\eqref{eq:def-mubar}, the differential equation \eqref{eq:eq_for_mubar}
can then be approximated as
\begin{equation}
\frac{d \overline{\mu}(t)}{dt}
\approx - 4\, \overline{\mu}(t)\;\overline{\gamma}(t)  \;
H(t)^2\,\tau_\text{QCD}
+ \text{O}\big(  H^3\,\tau_\text{QCD}^2 \big)\,,
\label{eq:eq_for_mubar_FRW}
\end{equation}
with a factor $|H|\,\tau_\text{QCD}$ in the first term
on the right-hand side from the nonanalytic
higher-derivative term in \eqref{eq:master-field-equation}
and a dimensionless function $\overline{\gamma}(t)$ from
the full master-field dynamics.
For comparison, analytic higher-derivative terms would give the
much smaller factor $H^2\,\tau_\text{QCD}^2$ contained in the second term
on the right-hand side of \eqref{eq:eq_for_mubar_FRW}.

The present Universe at coordinate time $t=t_0>0$
(setting the Big Bang coordinate time to zero, $t_\text{BB}=0$)
has a Hubble constant $H_0 \equiv H(t_0) \approx 1/t_0 >0$
and may be considered to have a vacuum state near
equilibrium, $\overline{\mu}=\mu_0+\delta\mu$, for $|\delta\mu/\mu_0|\ll 1$
and $\mu_0$ given by \eqref{eq:SSvacuum-mu0}.
The ordinary differential equation \eqref{eq:eq_for_mubar_FRW} gives then
approximately
\begin{equation}
\delta\mu
\approx -4\,\mu_0\;\overline{\gamma}(t_0)\;H_0^2\, t_0 \; \tau_\text{QCD}
\approx -\mu_0\;\gamma_0\;|H_0| \; \tau_\text{QCD}\,,
\label{eq:eq_for_mubar_deSitter}
\end{equation}
with all numerical factors absorbed in the constant $\gamma_0$
and using the positivity of $t_0$.
The chemical-potential shift \eqref{eq:eq_for_mubar_deSitter}
results in the following nonzero vacuum energy density \eqref{eq:rhovac(mu)}:
\begin{eqnarray}\label{eq:rhovac_present}
\rhovac(\mu)
&\approx&\epsilon_0- b_{1}\, q(\mu_0) - b_{1}\,(d q/d\mu)\,\delta\mu \nonumber\\[1mm]
&\approx&-   b_{1}\, \big(q_0^2\,\chi_0\big)\,\big(\delta\mu\big) \nonumber\\[1mm]
&\approx&   b_{1}\, \big(\Lambda_\text{QCD}^4/b_{1}\big)\,
                    \big(\mu_0\gamma_0\, |H_0| /\Lambda_\text{QCD}\big)
                    \nonumber\\[1mm]
&\approx&  \gamma_0\; \mu_0\;\Lambda_\text{QCD}^3\, |H_0|\nonumber\\[1mm]
&\approx&  \gamma_0\,b_1\; \Lambda_\text{QCD}^3\, |H_0|\,,
\end{eqnarray}
where the derivative of \eqref{eq:def-mubar} with respect to $q$
has been used in the second step,
the combined results \eqref{eq:SSvacuum-rhovac0}, \eqref{eq:compressibility},
\eqref{eq:scales}, and \eqref{eq:eq_for_mubar_deSitter} in the third step,
and \eqref{eq:SSvacuum-mu0} in the last step.
The final expression \eqref{eq:rhovac_present}, with positive
$\gamma_0\,b_1$ for a de-Sitter-like universe ($H_\text{deS} \approx H_0 >0$),
is precisely of the form \eqref{eq:Induced-rhovac} with  $f = \gamma_0\,b_1$.
Purely theoretically, the first two steps in \eqref{eq:rhovac_present}
highlight the importance of the vacuum compressibility
$\chi_0 \equiv \chi(q_0)$ for the dynamics of the vacuum energy density,
which has also been noted, for a different
model, in Eq.~(2.9) of Ref.~\cite{Klinkhamer2008}.

Result \eqref{eq:rhovac_present}
or equivalently \eqref{eq:Induced-rhovac} corresponds to a nonanalytic $|R|^{1/2}$
term in phenomenological $\widetilde{f}(R)$ modified-gravity theories,
where a tilde has been added to the function $f(R)$ in order
to distinguish it from the numerical factor $f$ used elsewhere in this
article (see Ref.~\cite{KlinkhamerVolovik2008c}
for references on this type of modified-gravity theories).
Specifically, the $\widetilde{f}(R)$ gravity induced by QCD is given by
\begin{equation}
\widetilde{f}(R)= - R - M\, \sqrt{|R|} + \cdots \,,
\label{eq:f(R)}
\end{equation}
with $M\geq 0$ and the same conventions for the Ricci scalar $R$ as in
Refs.~\cite{KlinkhamerVolovik2008b,KlinkhamerVolovik2008c,Weinberg2008}.
The $|R|^{1/2}$ term in \eqref{eq:f(R)} stands for all terms
$\square^n \big(|R|^{1/2}\,R^{-n}\big)$ with $n \in \mathbb{Z}$,
while the ellipsis indicates other higher-order terms in $R$.
Note that the complete gravitational action
may also have particular terms involving
the Ricci tensor and Riemann tensor, which,
for the de-Sitter metric, give a vanishing contribution to
the generalized Einstein field equation.

The modified-gravity model with $\widetilde{f}(R)$
from \eqref{eq:f(R)} belongs to the class of chameleon-type
models~\cite{KhouryWeltman2004,Faulkner-etal2007,Brax-etal2008}. For the case
of the gluon-condensate vacuum, the corresponding mass scale $M$ in
\eqref{eq:f(R)} is given by
\begin{eqnarray}
M &\sim& f\;\Lambda_\text{QCD}^3/E_\text{Planck}^2
\approx                                     
2 \times 10^{-34}\,\text{eV}\,\left(\frac{f}{0.004}\right)
\,\left(\frac{\Lambda_\text{QCD}}{200\,\text{MeV}}\right)^{3}\,,
\label{eq:f(R)_mass}
\end{eqnarray} 
which, up to a factor $4\pi$, corresponds to the inverse of \eqref{eq:t-cross-tmp}.
At small curvatures, $|R|\lesssim M^2$, the square-root term in \eqref{eq:f(R)}
becomes significant and leads to a large-distance
modification of gravity due to the existence of a gluon
condensate. It will be of interest to work out the
details of the corresponding cosmological model.

\vspace{2cm}  
\end{appendix}


\end{document}